\documentclass[aps,groupedaddress,longbibliography,notitlepage]{revtex4-1}

\usepackage{graphicx}
\usepackage{color}
\usepackage{amsmath}
\usepackage{amsfonts}
\usepackage{amssymb}
\usepackage{dcolumn}
\usepackage{hyperref}
\hypersetup{colorlinks=true,urlcolor=blue,linkcolor=blue,citecolor=blue}

\newcommand{\tfo}[4]{\ensuremath{{}_2F_1(#1 , #2 , #3 ; #4)}}

\begin{document}

\title{Numerical evaluation of the Gauss hypergeometric function: Implementation and application to Schramm-Loewner evolution}
\author{K.~J.~Schrenk}
\email{kjs73@cam.ac.uk}
\affiliation{Department of Chemistry, University of Cambridge, Lensfield Road, Cambridge, CB2 1EW, UK}
\author{J.~D.~Stevenson}
\email{js850@cam.ac.uk}
\affiliation{Department of Chemistry, University of Cambridge, Lensfield Road, Cambridge, CB2 1EW, UK}
\begin{abstract}
    Numerical studies of fractal curves in the plane often focus on subtle geometrical properties such as their left passage probability.
    Schramm-Loewner evolution (SLE) is a mathematical framework which makes explicit predictions for such features of curve ensembles.
    The SLE prediction for the left passage probability contains the Gauss hypergeometric function ${}_2F_1$.
    To perform computational SLE studies it is therefore necessary to have a method for numerical evaluation of ${}_2F_1$ in the relevant parameter regime.
    In some instances, commercial software provides suitable tools, but freely available implementations are rare and are usually unable to handle the parameter ranges needed for the left passage probability.
    We discuss different approaches to overcome this problem and also provide a ready-to-use implementation of one conceptually transparent method.
\end{abstract}
\maketitle
%
\section{Introduction}
Quantifying the geometry of fractal boundaries and paths has a long history in the study of phase transitions and lattice models.
In 2000, Schramm's work on loop-erased random walks introduced a new formalism devoted to this task \cite{Schramm00}.
Partially due to this formalism, called Schramm-Loewner evolution (SLE), it became possible to prove the existence and numerical values, e.g., of the 2D percolation critical exponents \cite{Smirnov01, Lawler01}.
Since SLE describes geometrical features of fractal curves, beyond the numerical value of the fractal dimension, it became a useful tool to understand several statistical physics models involving random paths and curves \cite{Cardy05, Bauer06, Gruzberg06, Henkel12b}.
Authoritative mathematical reviews of SLE can be found, e.g., in Refs.~\cite{Werner04, Lawler05, Rohde05, Smirnov06}.

Since SLE implies several geometrical properties of the associated curves, it is not clear \emph{a priori} whether a given distribution of fractal curves can in fact be described by SLE.
Therefore there have been a number of computer simulation studies testing numerically the compatibility of the properties of different ensembles of curves with the predictions of SLE \cite{Kennedy02, Bernard06, Amoruso06, Stevenson11, Stevenson11b, Daryaei12, Pose13}.
The present work is based on a note in Ref.~\cite{Schrenk14b}.

\begin{figure}
    \center
    \includegraphics[width=0.5\columnwidth]{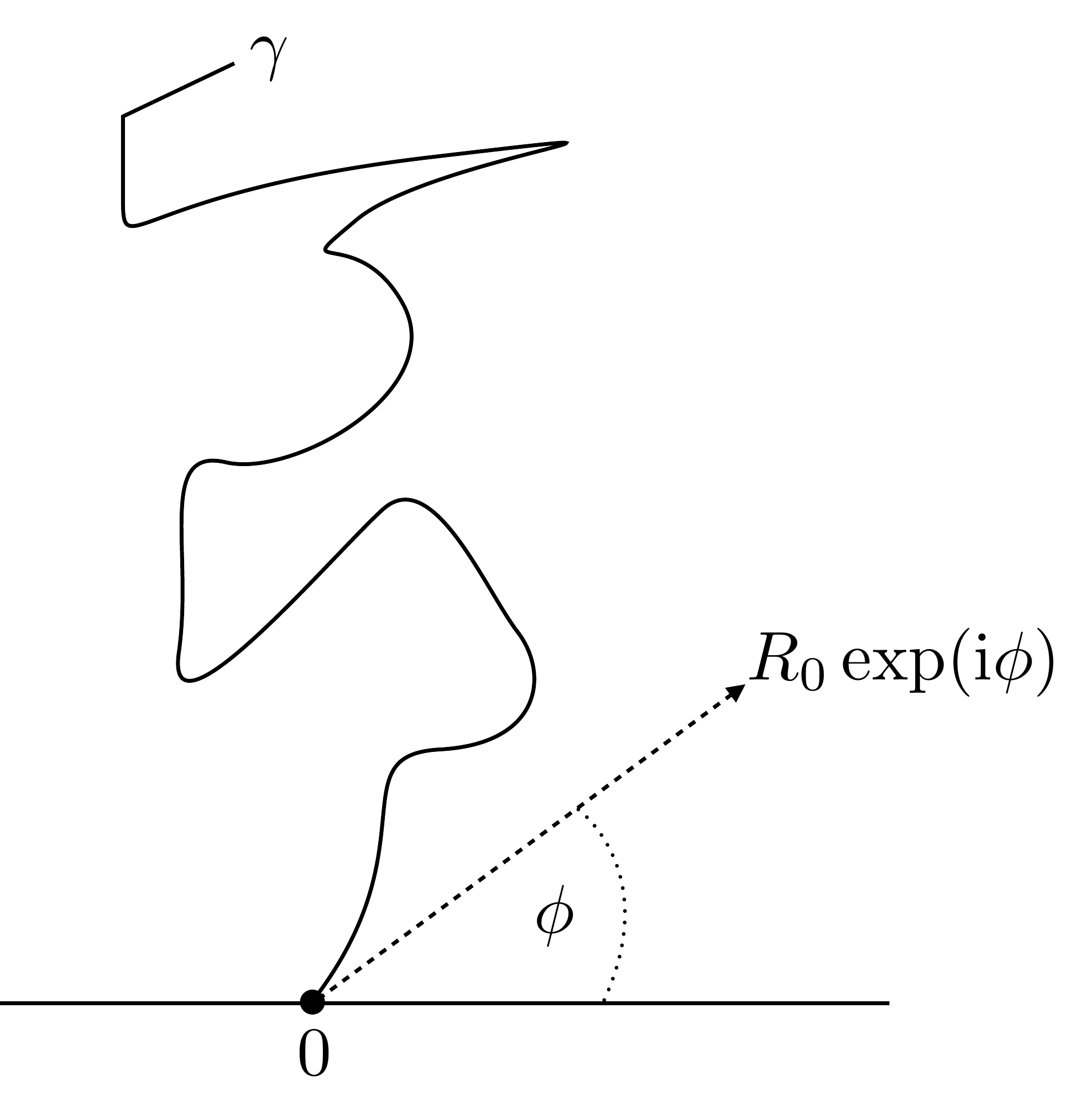}
    \caption{
        \label{fig::lpp_sketch}
        Sketch showing the geometrical definitions involved in the left passage probability formula in Eq.~(\ref{eqn::schramm_lpp_chord}).
        $P_\kappa(\phi)$ is the probability of a chordal SLE curve with parameter $\kappa$ to pass to the left of a point with argument $\phi$.
    }
\end{figure}

One of the most conclusive and subtle tests of SLE properties deals with the curves' left passage probability, in a certain geometrical setting (known as \emph{chordal SLE}).
For a curve originating in the origin and going to infinity in the upper half plane, Schramm proved \cite{Cardy92, Schramm01} that the probability $P_\kappa(\phi)$ that the curve passes to the left of the point
$z_0 = R_0 \exp(\text{i}\phi)$, see Fig.~\ref{fig::lpp_sketch}, is given by:
\begin{equation}
    \label{eqn::schramm_lpp_chord}
    P_{\kappa}(\phi)
    =
    \frac{1}{2} + \frac{\Gamma(4/\kappa)}{\sqrt{\pi}\Gamma(\frac{8-\kappa}{2\kappa})} \cot(\phi) \; {{}_2F_1}\left( \frac{1}{2}, \frac{4}{\kappa}, \frac{3}{2}; -\cot(\phi)^2 \right).
\end{equation}
Here, $\kappa \geq 0$ is the single parameter characterising the geometry of SLE curves [for Eq.~(\ref{eqn::schramm_lpp_chord}), $0 < \kappa < 8$] and known \cite{Lawler05} to be related to the curves' fractal dimension $d_f$ by
\begin{equation}
    \label{eqn::df_kappa_rel}
d_f = \min\{1 + \kappa/8,\ 2\}.
\end{equation}

\begin{figure}
    \center
    \includegraphics[width=0.8\columnwidth]{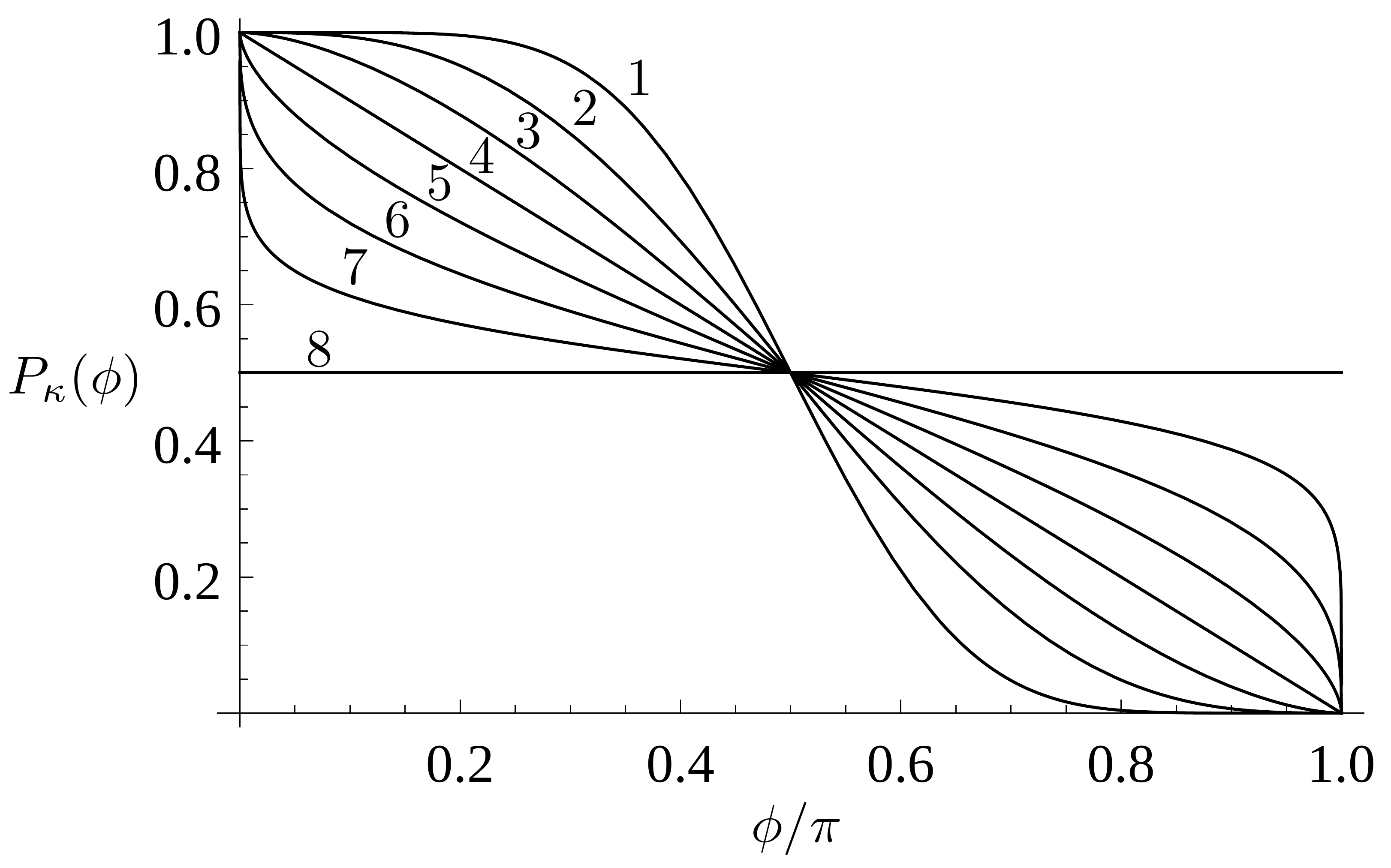}
    \caption{\label{fig::schramm_formual_eval_mathematica}
        Graph of Eq.~(\ref{eqn::schramm_lpp_chord}):
        For different values of $\kappa$, $P_\kappa(\phi)$ is shown versus the angle $\phi/\pi$.
        The corresponding $\kappa$ values are shown next to the curves.
        Note that, e.g.,
        $P_8(\phi) = 1/2$,
        $P_4(\phi) = 1-\phi/\pi$,
        and
        $P_2(\phi) = 1 + [\sin(2\phi)/2-\phi]/\pi$
        \cite{Schramm01}.}
\end{figure}

As seen from Eq.~(\ref{eqn::df_kappa_rel}),
$d_f = 1$
corresponds to
$\kappa = 0$
and
$d_f = 2$
is recovered for
$\kappa \geq 8$.
We note that the left passage probability $P_\kappa(\phi)$ of an ensemble of curves depends only on $\kappa$ and the angle with the real axis $\phi$.
Figure \ref{fig::schramm_formual_eval_mathematica} shows $P_\kappa(\phi)$ as function of $\phi$, for different values of $\kappa$.
For
$\phi = \pi / 2$,
one has
$P_\kappa(\pi / 2) = 1/2$,
independent of $\kappa$,
as expected by symmetry.
For
$\kappa = 8$,
the curves are space-filling and 
$P_8(\phi) = 1/2$,
independent of $\phi$ \cite{Schramm01}.

The data in Fig.~\ref{fig::schramm_formual_eval_mathematica} has been obtained by evaluating Eq.~(\ref{eqn::schramm_lpp_chord}) with Mathematica \cite{Mathematica}.
However, in numerical tests of SLE predictions, one needs to evaluate Eq.~(\ref{eqn::schramm_lpp_chord}) many times \cite{Chatelain10, Saberi10, Norrenbrock12, Daryaei13b} and it can be advantageous to do the evaluations within the simulation run.
In many cases, corresponding simulations are performed using the C++ programming language and the associated libraries \cite{Stroustrup09}.
The required numerical evaluations of the cotangent and gamma \footnote{We also tested  evaluating $\Gamma$ by integrating the appropriate contour integral, as proposed in \cite{Schwartz12}, obtaining satisfactory results.} functions can be handled by the C++ standard library, the gnu scientific library (GSL) \cite{gsl} and the Boost libraries \cite{boost}.
Although GSL and Boost, provide implementations of ${}_2F_1$ these do not cover the parameter regimes required for the left passage probability.

The remainder of this work is structured as follows.
In Sec.~\ref{sec::methods}, we describe different approaches of how to numerically evaluate ${}_2F_1$ as needed for the left passage probability.
Use of argument transformations is reviewed in Sec.~\ref{ss:argtrans} and integrating ${}_2F_1$'s integral representation is discussed in Sec.~\ref{sec::intmethod}.
We draw conclusions in Sec.~\ref{sec::conclusion}.
\section{\label{sec::methods}Methods}
We note that the evaluation of ${}_2F_1$ is in general a difficult numerical task requiring a sophisticated combination of various techniques \cite{Pearson09, Forrey97, Michel08}.
Here we focus on methods useful for the parameter regime relevant to the left passage probability formula in Eq.~(\ref{eqn::schramm_lpp_chord}).
\subsection{\label{ss:argtrans}Argument transformations}
Many implementations of ${}_2F_1$, such as those in the GSL, Boost, or Cephes \cite{cephes} libraries
only support arguments $\vert z \vert < 1$.  However, for the  left passage probability in Eq.~(\ref{eqn::schramm_lpp_chord}), 
$z = -\cot(\phi)^2$, therefore $z$ can in general take any non-positive value.
To use such an implementation for $\vert z \vert \geq 1$, known relations can be applied to transform the 
problem back to the unit disk \cite{Forrey97, Michel08}.  An implementation
in a large public library which includes these transformations is SciPy \cite{Scipy2F1}.  Though the SciPy interface is in python, the underlying algorithm
is in c and was adapted from the Cephes mathematical library \cite{cephes} to support $z < -1$.

For $-\infty < z < -1$, Ref.~\cite{Forrey97}, proposes to use the following transformation (Abramowitz and Stegun (A\&S) \cite{Oberhettinger72} eq. $15.3.8$):
\begin{eqnarray}
    \tfo{a}{b}{c}{z}
    &=& (1-z)^{-a} \frac{\Gamma(c)\Gamma(b-a)}{\Gamma(b)\Gamma(c-a)} \tfo{a}{c-b}{a-b+1}{(1-z)^{-1}}\\
    &+& (1-z)^{-b} \frac{\Gamma(c)\Gamma(a-b)}{\Gamma(a)\Gamma(c-b)} \tfo{b}{c-a}{b-a+1}{(1-z)^{-1}}.
\end{eqnarray}
SciPy, which we consider for the comparison in Fig.~\ref{fig::cpp_result_errpr}(b), uses, e.g., for $-\infty < z < -2$, the following relation (A\&S \cite{Oberhettinger72} eq. $15.3.7$):
\begin{eqnarray}
    \tfo{a}{b}{c}{z}
    &=&
    (-z)^{-a}
    \frac{\Gamma(c)\Gamma(b-a)}{\Gamma(b)\Gamma(c-a)}
    \tfo{a}{1-c+a}{1-b+a}{1/z}\\
    &+&
    (-z)^{-b}
    \frac{\Gamma(c)\Gamma(a-b)}{\Gamma(a)\Gamma(c-b)}
    \tfo{b}{1-c+b}{1-a+b}{1/z}.
\end{eqnarray}
When using these transformations care must be taken to avoid special parameter 
values that are numerically difficult.  For example, when $b-a$ is an integer 
(which will happen for $\kappa = 8/3$) $\Gamma(b-a)$ or $\Gamma(a-b)$ will diverge.  In these cases, or
when $z$ is close to $-1$, one could use the transformations A\&S $15.3.4$ or A\&S $15.3.5$ \cite{Oberhettinger72, Scipy2F1}.
\subsection{\label{sec::intmethod}Integral representation}
As an alternative to using argument transformations, we next apply the method of special function evaluation proposed by Schwartz \cite{Schwartz69, Schwartz12, Schwartz13}.
${}_2F_1$ can be expressed as
\begin{equation}
    \label{eqn::gauss_hyper_integral}
    {}_2F_1(a,b,c;z) = \frac{\Gamma(c)}{\Gamma(b)\Gamma(c-b)} \int_0^1 t^{b-1}(1-t)^{c-b-1}(1-zt)^{-a} \text{d}t,
\end{equation}
where ${\text{Re}(c)>\text{Re}(b)>0}$, ${\vert\arg(1-z)\vert<\pi}$, and $\Gamma$ is the gamma function \cite{Kratzer63, WeissteinHypergeometric, FunctionsWolfram2F1, NISTFunctions}.
For computing numerically the integral in Eq.~(\ref{eqn::gauss_hyper_integral}) we use the method proposed in Refs.~\cite{Schwartz12, Schwartz13}.
The main point of this method is to look at integral representations of special functions.
Where necessary, they are reformulated as infinite integrals over quickly decaying integrands.
These integrals are then computed with the trapezoidal rule:
\begin{equation}
    \label{eqn::schwartz_method}
    \int_{-\infty}^\infty f(x) \text{d}x \approx \sum_{k} hf(kh),
\end{equation}
where $h$ denotes the step size.
In computational practice, after reaching a desired accuracy, the sum in Eq.~(\ref{eqn::schwartz_method}) is truncated (see Ref.~\cite{Schwartz12}, and for a recent mathematical treatment see Ref.~\cite{Trefethen13}).
This allows to move the singularities, for ${t=0}$ and ${t=1}$, of the integrand in Eq.~(\ref{eqn::gauss_hyper_integral}) to infinity.
Therefore they do not affect the truncated sum of Eq.~(\ref{eqn::schwartz_method}) \cite{Schwartz13}.
For example, this can be achieved by a changing the integration variable (from $t$ to $u$) as:
\begin{equation}
    t = \frac{1+\tanh[\sinh(u)]}{2}.
\end{equation}

To use the integral representation of Eq.~(\ref{eqn::gauss_hyper_integral}) in practice, some algebraic operations need to be performed.
Looking at the expression for the left passage probability in Eq.~(\ref{eqn::schramm_lpp_chord}), one sees that for our purpose we only need to evaluate 
${}_2F_1(a,b,c;z)$ with real parameters
$a=1/2$,
$b=\kappa/4$,
and
$c=3/2$.
The argument $z = -\cot(\phi)^2$ is also real.
To obtain parameter values that are suitable for the numerical integration of the integral representation, we apply the following combination of known results \cite{Kratzer63, WeissteinHypergeometric} for ${}_2F_1$:

To evaluate $\tfo{a}{b}{a}{z}$, we use
\begin{equation}
    \tfo{a}{b}{a}{z} = (1-z)^{-b}.
\end{equation}

To evaluate $\tfo{a}{b}{b}{z}$, we use
\begin{equation}
    \tfo{a}{b}{b}{z} = (1-z)^{-a}.
\end{equation}

Since we have $a = 1/2 >0$, we apply the following transformation, based on the recursion relations for the Gauss hypergeometric function, to lower the value of $a$:
\begin{eqnarray}
    \tfo{a}{b}{c}{z}
    &=& \frac{a-1-c}{(a-1)(z-1)}\tfo{a-2}{b}{c}{z}\\
    &+& \frac{c-2(a-1)+(a-1-b)z}{(a-1)(z-1)}\tfo{a-1}{b}{c}{z}.
\end{eqnarray} 

In case $\kappa < 4$, we have $b < 1$, and apply the following recursion relation to increase the value of $b$ in the integral:
\begin{eqnarray}
    \tfo{a}{b}{c}{z}
    &=& \frac{(b+1)(z-1)}{b+1-c}\tfo{a}{b+2}{c}{z}\\
    &-& \frac{c-2(b+1)+(b+1-a)z}{b+1-c}\tfo{a}{b+1}{c}{z}.
    \label{eqn::tfo_recursion_b}
\end{eqnarray}

Using Eq.~(\ref{eqn::tfo_recursion_b}), we can end up with $c \leq b$; in this case, we apply 
\begin{eqnarray}
    \tfo{a}{b}{c}{z}
    &=& \frac{(a-c-1)(b-c-1)}{c(c+1)(1-z)}\tfo{a}{b}{c+2}{z}\\
    &-& \frac{-c+(2c+1-a-b)z}{c(1-z)}\tfo{a}{b}{c+1}{z},
\end{eqnarray}
raising the value of $c$.
\begin{figure}
    \center
    (a)
    \includegraphics[width=0.8\columnwidth]{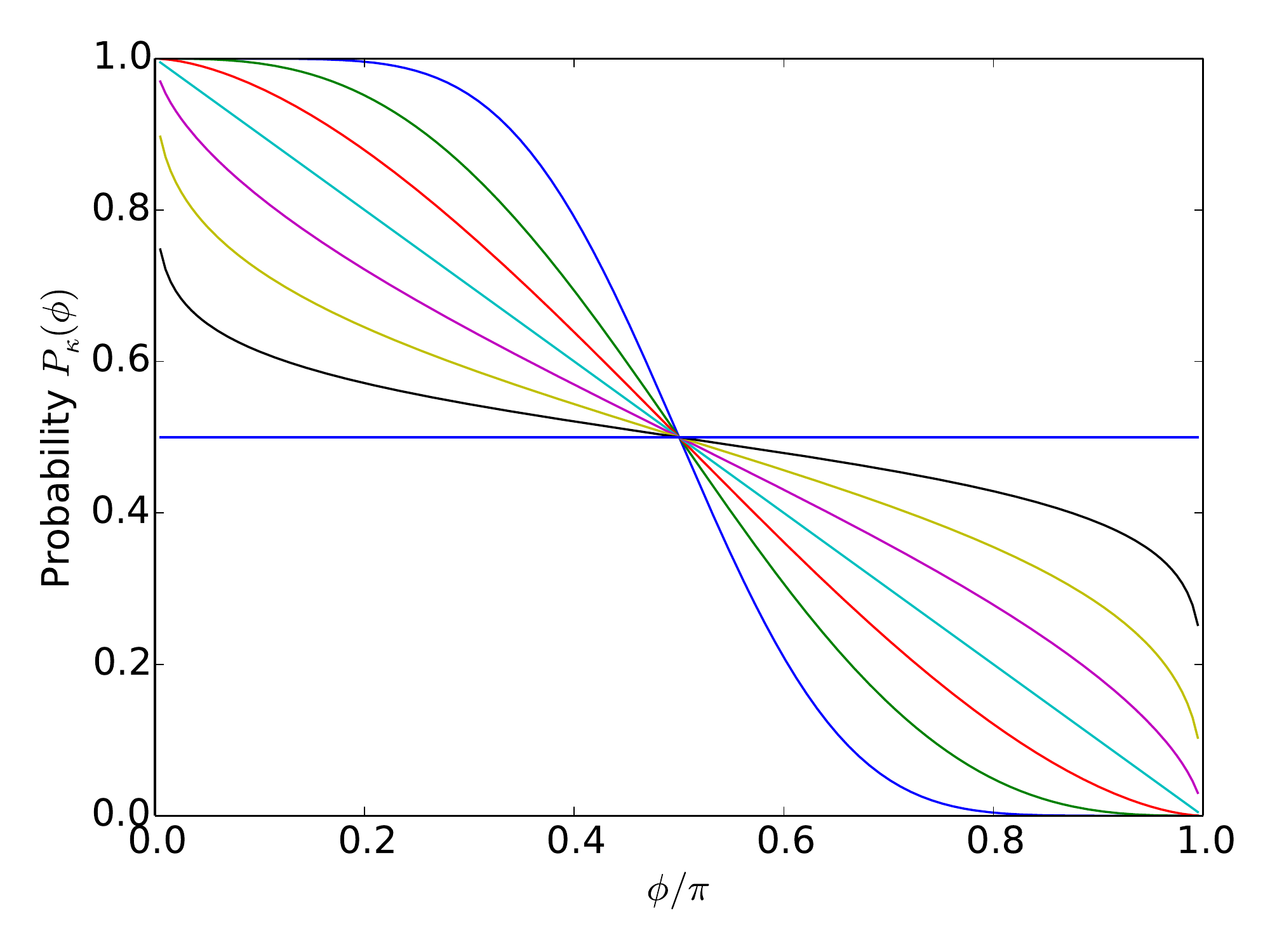}
    (b)
    \includegraphics[width=0.8\columnwidth]{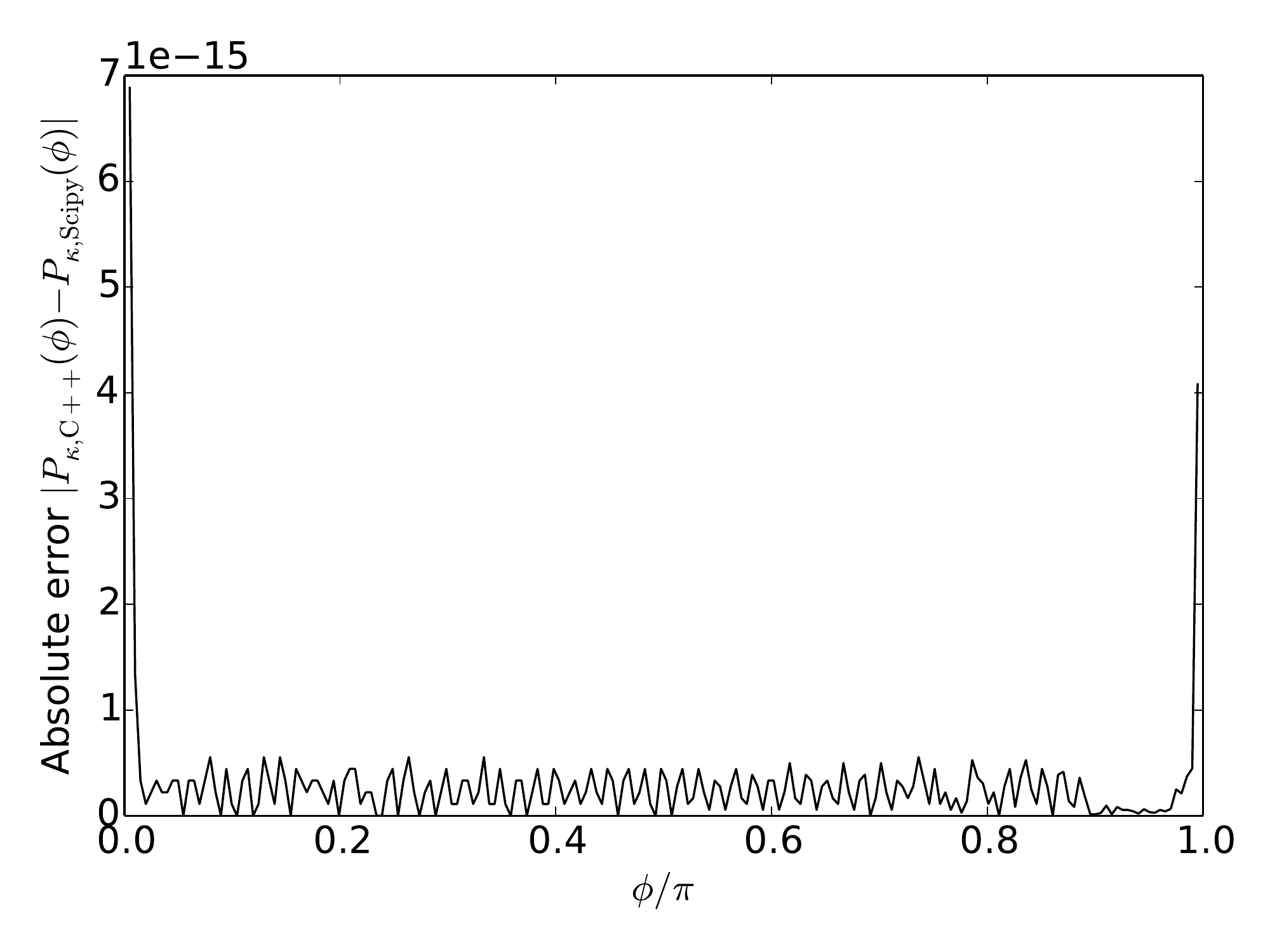}
    \caption{\label{fig::cpp_result_errpr}
    (a) Left-passage probability $P_\kappa(\phi)$ as function of $\phi$, evaluated using the method described in this work.
    For $\kappa = 8$, the curve is horizontal and it becomes more inclined with increasing $\kappa$.
    The considered $\kappa$ values range from $1$ to $8$ as in Fig.~\ref{fig::schramm_formual_eval_mathematica}.
    (b) Absolute value of the difference between $P_4(\phi)$, as described in section \ref{sec::intmethod}, and a control method using the Scipy (Python) ${}_2F_1$ implementation, as function of the angle $\phi$.
    }
\end{figure}

The method for evaluating the left passage probability based on the integral representation has been implemented in C++ and is available online \cite{LPPRepo}.
The class \texttt{schramm$\_$equation} located in the header file \texttt{schramm$\_$equation.hpp} takes the value of $\kappa$ in the constructor and provides a bracket operator which returns the numerical value of $P_\kappa(\phi)$.
Application of the recursion relations and processing of special parameter values, as discussed above, is handled by recursive function calls in the header \texttt{gauss$\_$transformations.hpp}.
This, in turn, calls the class \texttt{gauss$\_$hypergeometric} which performs the numerical integration.

Figure \ref{fig::cpp_result_errpr}(a) shows the left-passage probability as function of $\phi$, for different values of kappa, as evaluated by the method discussed in this section.
To test our numerical evaluation of the left passage probability, we compare the results obtained by our C++ program to results obtained using the Scipy (Python) \cite{Scipy2F1} implementation of ${}_2F_1$, which uses the argument transformations described in Sec.~\ref{ss:argtrans}.
An example of this comparison is shown in Fig.~\ref{fig::cpp_result_errpr}(b) which plots the absolute difference between $P_4(\phi)$, as obtained from the presented method and from the Python library function, as function of $\phi$.
\section{\label{sec::conclusion}Conclusion}
We discussed methods for the numerical evaluation of the Gauss hypergeometric function ${}_2F_1$.
While the presented options seem to be satisfactory for the parameter regime relevant for the evaluation of the Schramm-Loewner evolution (SLE) left passage probability, in general the numerical evaluation of ${}_2F_1$ is a complicated task and subject of ongoing research \cite{Pearson09}.
Hopefully, our work will facilitate numerical studies of SLE in the future.
Given its importance for science, it would be useful if authoritative C++ libraries, such as the standard library \cite{Stroustrup09}, Boost \cite{boost}, 
or GSL \cite{gsl}, provided implementations of ${}_2F_1$ that covered the relevant parameter regimes.
\section*{Acknowledgments}
K.J.S. acknowledges useful discussions with Malte Henkel (Universit\'e de Lorraine Nancy) and Nicolas Pos\'e (ETH Zurich), as well as support by the Swiss National Science Foundation under Grant No. P2EZP2-152188.

\end{document}